# The status of the search for low mass WIMPs: 2012

David B. Cline
Astroparticle Physics
UCLA

**Abstract**

Using information from a recent dark matter symposium at Marina del Rey, we discuss the most recent evidence and constraints on low mass WIMPs. There are now five separate experimental limits on such WIMPs, including a new paper on the XENON100 225 day exposure. There are very different experimental methods with different backgrounds that comprise this limit. We speculate on the possible sources of the reported low mass WIMP signals and background.

**Introduction**

With the discovery of a 125 GeV particle by CMS [1] and Atlas [2] that is widely believed to be the Higgs boson, various models of supersymmetric WIMPs increase the expected mass to the 500 GeV or greater and cross-sections to between $10^{-45}$ to $10^{-47}$ cm$^2$ . Only the lower edge of this region has been explored by XENON 100s: 225 day exposure [3]. Our previous work described the search for low mass WIMPs [4].

The likelihood of a supersymmetric low mass WIMP from the theory is very remote. Nevertheless claims from DAMA, CoGeNT, and CRESST have not been withdrawn. This is an unfortunate problem in the worldwide search for dark matter particles. At the recent Dark Matter symposium at Marina del Rey there was very strong evidence put forth to limit the possibility of low mass WIMPs [5]. In particular the null CDMS II search for annual variation in the low mass range coupled with the latest XENON100, 225 day exposure strongly constrains the low mass WIMP hypothesis [6].

In this paper we will present all the current evidence for the low mass WIMP search.

## 1. Summary of world limit on low mass WIMP signals

In Figure 1 we show a summary of the current limits on the low mass WIMP region [3]. We note that the CDMS II, Simple, XENON10 limits come from very different methods:

- CDMS II            Ultra cold Ge detector

- XENON10     Use of the S2 only signal from a special low threshold run of XENON 10

- SIMPLE       Use of heated droplets

- XENON100    Use of Xenon detector using (2 experiments) traditional methods of $S_1$ and $S_2$

Because the claimed cross-section is so large, these methods are all very robust.

The limits from XENON100 deserve a special discussion [3]. Both the 100 day XENON100 exposure and the more recent 225 day exposure are inconsistent with a low mass WIMP to the 90 percent confidence level. These data are totally independent and not summed in Figure 1. One could assume that the new 225-day data logically reinforce the 100-day limit. There are then five limits: Simple, XENON10 (S2), CDMS II, XENON100, 100 days, and XENON 225-day limits. All are independent and are 90 percent confidence level null limits. We note that the DAMA results are reported as $3\sigma$ limits (see Reference 4 for references).

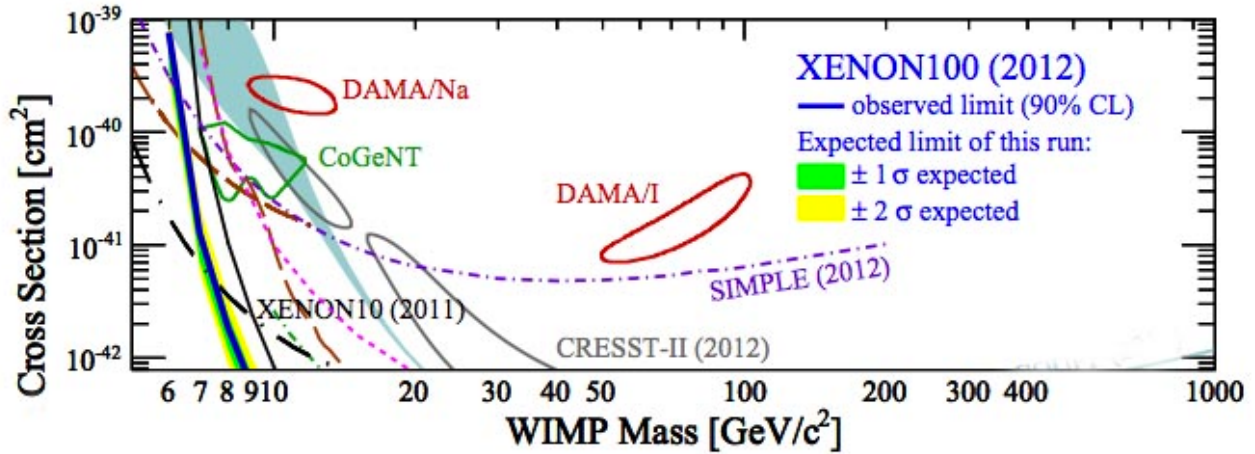

Figure 1. An enlargement of the low mass scale of WIMP searches from the recent XENON100 225 day paper (E. Aprile et al, "Dark Matter Results from 225 Live Days of XENON100 Data," http://arxiv.org/abs/1207.5988).

## 2. The use of the $S_2$ signal from a Xenon on Argon dark matter detector

Normally the $S_2$ signal is used to carry out discrimination of a WIMP signal is from an EM background. However, as shown by the ZEPLIN II group who first measured $S_2$ experimentally [7] and from experiments by the UCLA Torino team in the 1990s, this parameter is very robust. When a particle hits a large atom like Xenon the outer electrons are easily stripped off. With an electric field applied the free electrons drift to a gas system that provides amplification. Typically one electron from the vertex can yield 20 to 30 photoelectrons in the experiment's PMTs, giving $S_2$. $S_2$ is usually used to trigger the detector. It was recognized early on that the $S_2$ signal could be used to measure energy in low energy events [8]. Recently a UCLA study has shown a way to analyze data using the $S_2$ signal [9]. See also the work of P. Sorensen and the XENON10 group [10]. The essence of this subject is that very low energy recoils can be detected with the $S_2$ signal, making this appropriate to search for very low mass WIMP signals. The $S_2$ signal has a lower threshold than any other current dark matter detector.

There are two choices:

(a) Use a small $S_1$ signal to determine the position of the event in the detector and use $S_2$ to measure the energy [9];

(b) Use the diffusion of the $S_2$ signal to measure Z (upward) position and determine x, y (z) for the event [10].

The (b) method has been used to determine the limit on low mass WIMPs shown in Figure 2 showing a very robust limit (also shown in Figure 1) [10].

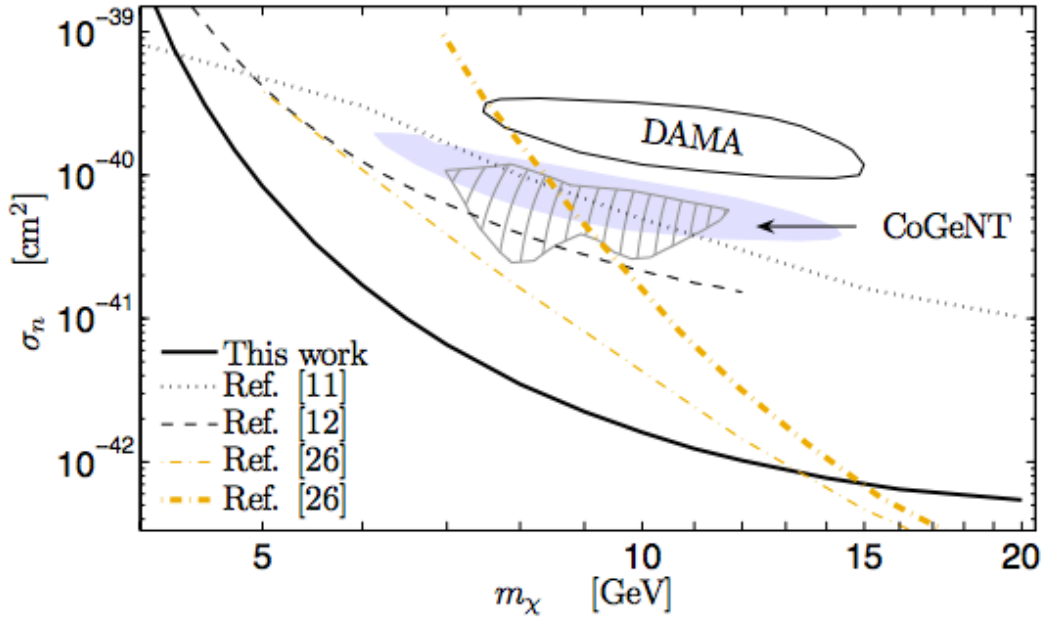

Figure 2. Curves indicate 90% C.L. exclusion limits on spin-independent $\sigma_n$ for elastic dark matter scattering, obtained by CDMS (dotted [11], and dashed [12]) and XENON100 (dash-dot [26]). The region consistent with assumption of a positive detection by CoGeNT is shown (hatched) [2], and (shaded)[4]; the latter assumes a 30% exponential background. Also shown is the 3σ allowed region for the DAMA annual modulation signal (solid contour) [40] (see Reference 10).

## 3. The Search for Annual Signal Variation with the CDMS II

This CDMS II result is remarkable since not even the expected Radon annual variation is observed. Once a real WIMP signal is observed the observation of annual signal variation is a powerful method to prove that WIMPs have been discovered. At the recent Marina Del Rey Dark Matter conference the CDMS II group presented a search for the Annual Signal Variation observed by the Cogent experiment (recall that CDMS II is a 5kg detector and Cogent is 300 grams). In Figure 3 we show the CDMS II results.

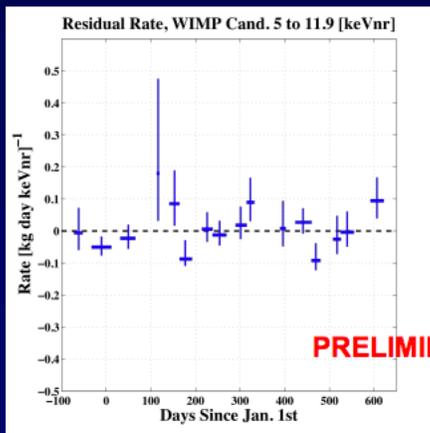
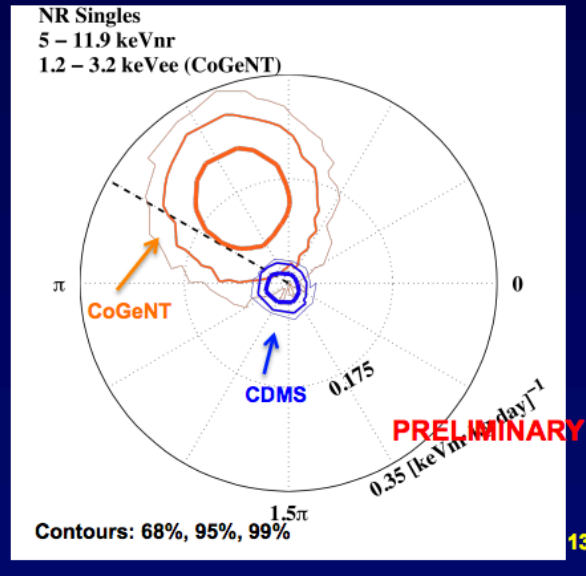

Figure 3. The CDMS group has searched for a low energy signal using the low noise components of the detector as shown in Figure 4. These limits are also shown in Figure 4 (from the CDMS II talk in Reference 5).

In Figure 4 we show the current limits from CDMS II on the low mass region. This analysis used low noise sensors on CDMS II to set this impressive limit [6]. These limits are also shown in Figure 1 and exclude even an enlarged region for DAMA and CoGeNT signals.

### 4. Neutron signals underground

It is well known that the neutron flux in underground labs has an annual variation. This is likely due to the amount of water or snow in the over burden. In the winter the water absorbs neutrons, in the summer much less so. The ICARUS group measured the LGNS neutron flux as shown in Figure 5. Note that this annual variation fits the DAMA data. DAMA is also at the LGNS. J. Ralston took the ICARUS results and extrapolated over the entire DAMA region (Figure 5) (this is not a fit). Note the excellent agreement with the data. We are not claiming that neutrons make the signal in DAMA, only that there are underground sources that seem to fit the same annual variations than one not due to WIMPs.

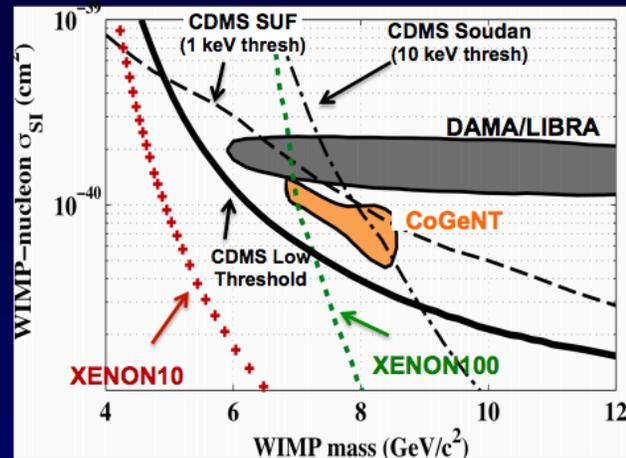
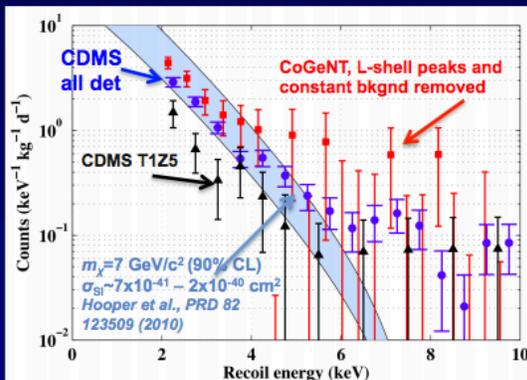

Figure 4. Limits on the low mass WIMP signal from a presentation by the CDMS II group at the Marina del Rey symposium February 2012.

## 5. Signals of Annual Variation underground and DAMA

There are several processes that cause annual variation of processes underground that are similar to the DAMA results.

1. *Radon abundance*
   -Has a clear annual increase in the summer and decrease in the winter seen in all underground laboratories
2. *Variation of neutron flux*
   -In Figure 5 we show neutron intensity data from ICARUS expanded and compared with the DAMA results. All underground laboratories see a neutron flux annual variation.
3. *The annual variation of cosmic muons as compared with DAMA data (Figure 3)*
   -In Figure 6 we show the LVD muon data and compare with the DAMA results (as discussed in Section 4). We do not claim a good fit but there is a general agreement.

For all we know DAMA may be seeing a combination of such effects and the phase they observe would be a mixture of these events. Until we identify the actual source of the signals we will not know the actual phase to predict.

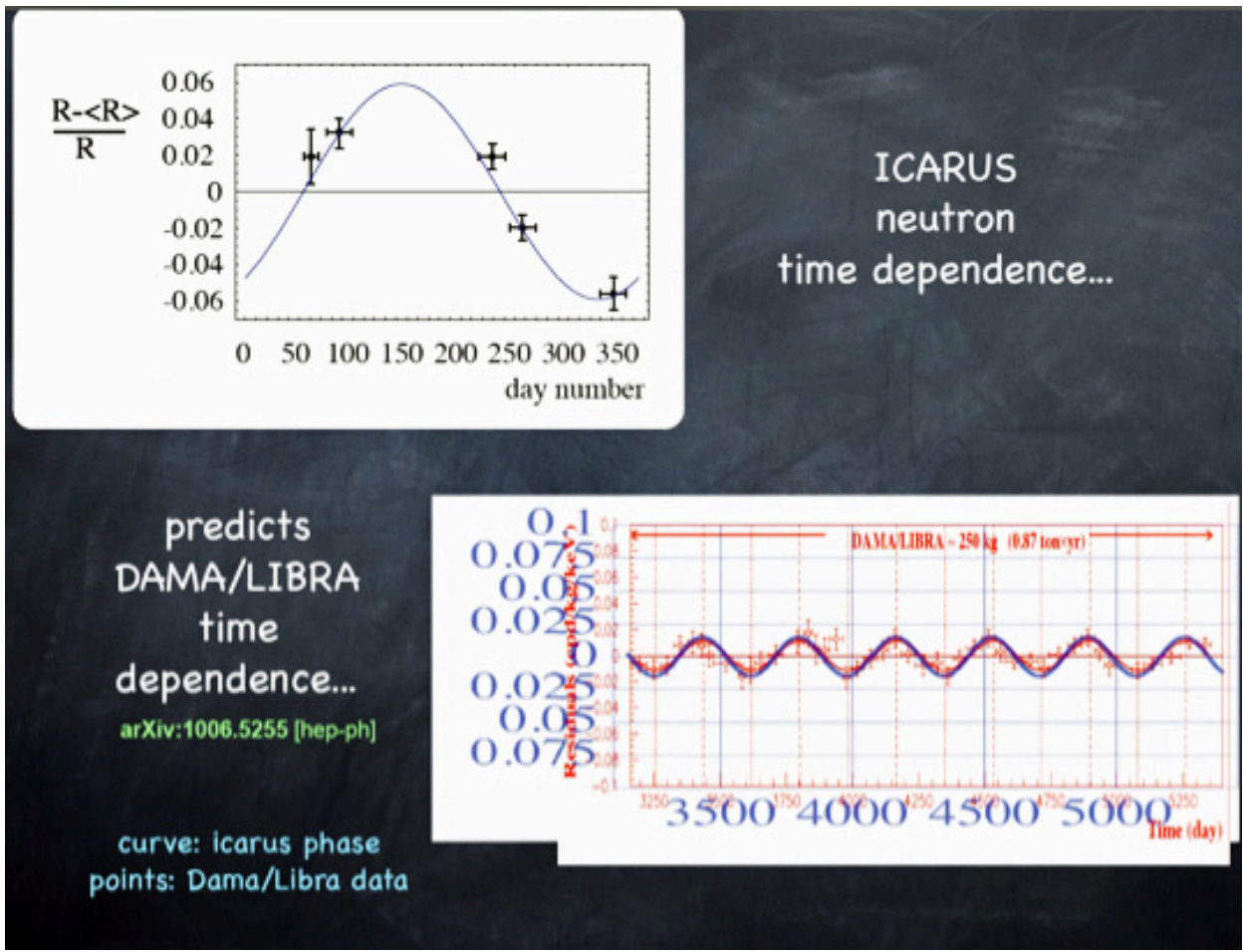

Figure 5. A study of neutron events at the LNGS by the ICARUS group extrapolated to the DAMA results by Ralston (arXiv 1006.5255).

## 6. Conclusion

The search for low mass dark matter has been mostly negative. While several signals (DAMA, CoGeNT, CRESS II) suggest low mass WIMPs, there are very strong experimental constraints on these signals.

1. 5 separate experiments do not see a nuclear recoil signal consistent with a low mass WIMP (Figure 1).
2. The specific search by CDMS II for annual variation or a direct signal for either CoGeNT or DAMA is null even if the DAMA region is greatly expanded (Figures 3 and 4).
3. While not discussed in this paper, the direct comparison of the singles rates in DAMA with a carefully determined radioactive background finds no evidence for a WIMP excess on the data [1]. A similar study has been carried out by Peter Smith at UCLA (unpublished).
4. There are several experimental sources of annual variation background that can cause signals in underground detectors. The neutron background measured by ICARUS seems to give a similar signal but others such as Radon and muon also give annual variation. These backgrounds are observed at all underground laboratories and have a simple explanation such as the water load charges in the overburden or the change in density of the upper atmosphere.

We thank the participants of the Marina del Rey Dark Matter and Dark Energy meeting February 2012 and members of the XENON 2012 collaboration.

This paper was prepared at the Aspen Center for Physics, Summer 2012. We thank this center for help.

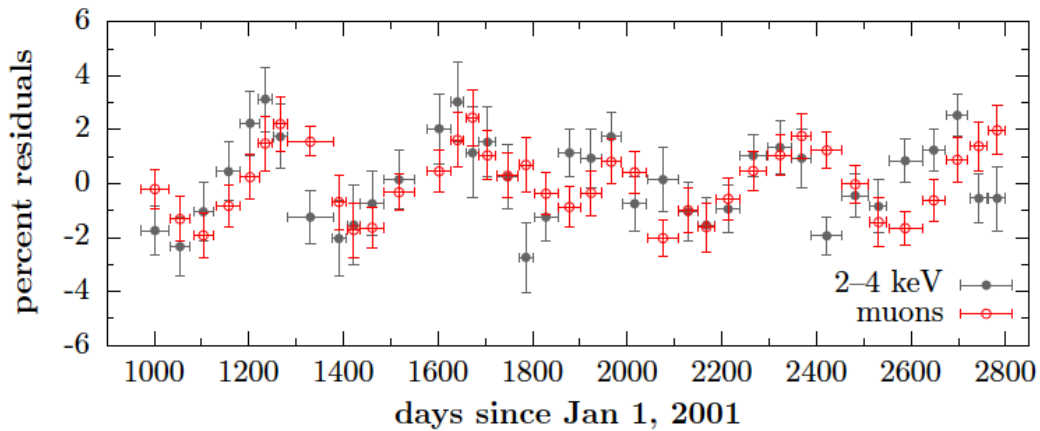

Figure 6. A comparison of the LVD muon data and the DAMA results shown at the Marina del Rey Dark Matter symposium.